\documentclass[aps,pra,twocolumn,groupedaddress]{revtex4}
  \advance\oddsidemargin  by -1in
  \advance\evensidemargin by -1in
\begin{document}
\newcommand{\non}{\nonumber}
\newcommand{\be}{\begin{equation}}
\newcommand{\ee}{\end{equation}}
\newcommand{\bq}{\begin{eqnarray}}
\newcommand{\eq}{\end{eqnarray}}
\title{
Collective modes, chaotic behavior and self-trapping
in the dynamics of three coupled Bose-Einstein condensates}
\author{Roberto Franzosi$^{*}$ and Vittorio Penna$^{\dagger}$}
\affiliation{$^* \!$ Dipartimento di Fisica 
dell'Universit\`a di Pisa $\&$ INFN, Sezione di Pisa, 
Via Buonarroti 2, I-56127 Pisa, Italy.\\ }
\affiliation{$^{\dagger} \!$ Dipartimento di Fisica $\&$ UdR INFM, 
Torino Politecnico,
%and INFM, Unit\`a di Ricerca di Torino,
C.so Duca degli Abruzzi 24, I-10129 Torino, Italy.}
\date{\today}
%%%%%%%%%%%%%%%%%%%%%%%%%%%%%%%%%%%%%%%%%%%%%%%%%%%%%%%%%%%%%%%%%%
\begin{abstract}
The dynamics of the three coupled bosonic wells (trimer)
containing $N$ bosons is investigated within a standard
(mean-field) semiclassical picture based on the coherent-state
method.
Various periodic solutions (configured as $\pi$-like, dimerlike
and vortex states) representing collective modes are obtained
analitically when the fixed points of trimer dynamics are
identified on the $N$=const submanifold in the phase space. 
Hyperbolic, maximum and minimum points are recognized in the
fixed-point set by studying the Hessian signature of the
trimer Hamiltonian.

The system dynamics in the neighbourhood of periodic
orbits (associated to fixed points) is studied via numeric
integration of trimer motion equations thus revealing
a diffused chaotic behavior (not excluding the presence of
regular orbits), macroscopic effects of population-inversion
and self-trapping.
In particular, the behavior of orbits with initial conditions
close to the dimerlike periodic orbits shows how the self-trapping
effect of dimerlike integrable subregimes is destroyed by the
presence of chaos.

\medskip
\centerline{PACS numbers: 03.75.Fi, 05.45.-a, 03.65.Sq}
\end{abstract}
\maketitle
%%%%%%%%%%%%%%%%%%%%%%%%%%%%%%%%%%%%%%%%%%%%%%%%%%%
%\pacs{PACS XX, YY, ZZ}
%N. 74.20.-z, 05.30.Fk}
%
%%%%%\begin{multicols}{2}
%%%%%%%%%%%%%%%%%%%%%%%%%%%%%%%%%%%%%%%%%%%%%%%%%%%%%%%%%%%%%%%%%%%%%%%
\section{Introduction}
%
%\bigskip

%Many progress
A remarkable progress 
in the experimental design has
been done since the first direct
observation of Bose-Einstein condensation in diluite
atomic gas \cite{PEXP}.
One of most promising development concerns
the construction of experimental devices in which condensates,
achieved within complex geometries, interact with each other
giving rise to quantum effects that are observable at the
macroscopic level~\cite{COLL} -\cite{INGU}.
In this respect, one should recall,
for example, the (superfluid) boson Josephson-junction arrays
obtained by means of optical lattices that trap weak interacting
BECs in periodic arrays of potential wells \cite{JJC}.
In parallel with the experimental work,
an increasing attention has been devoted theoretically
to study the dynamical behavior of low-energy states
%weakly excited states
in arrays of Bose-Einstein condensates (BEC) where 
the number of lattice sites (namely the potential wells occupied
by the condensate) is very large~\cite{LTR,TROMB}.

Opposite situations, corresponding to
`lattices' formed by two or three interacting wells,
have been investigated as well in various recent
papers (see Refs. \cite{MILB} -\cite{FP2}). In particular,
the two-well system (dimer) has been analysed thorougly
from both the semiclassical (mean-field) and
the purely quantum viewpoint in
Refs. \cite{SM1,SM2} and \cite{FPZ1,MACZ,FP2},
respectively. Such investigations have revealed
how the nontrivial structure of dimer phase space
causes many significant phenomena such as the
symmetry-breaking effect (issuing oscillation
modes that are isoenergetic but inequivalent),
the onset of $\pi$-phase oscillations,
the {\it self-trapping}
of boson populations, and, quantum-mechanically,
the occurrence of (parameter-dependent, nondegenerate)
doublets in the energy spectrum entailing periodic
self-trapping.

The connection between the quantum and the
semiclassical picture of many interacting bosonic wells
has been illustrated in Ref. \cite{FPZ1} and, in view of the 
closed link between an array of interacting BECs and the 
Bose-Hubbard (BH) model~\cite{FPZ1,CIR}, in Ref. \cite{AP} 
within the BH model theory.
We wish to observe that the semiclassical approach
(corresponding to describing condensates within the
Bogoliubov approximation) is appropriate for interacting
wells with macroscopic boson populations.
However, the systems recently obtained, where condensate is
distributed among many wells are able to provide mesoscopic
numbers of bosons per wells~\cite{INGU}, might be best
modeled by using the space-mode approximation~\cite{MILB,PW}
stemmed from the second-quantized boson field theory.
%of boson fluids.

Compared with the dimer nonlinear behavior,
indeed the system of three interacting bosonic wells
({\it trimer model}) exhibits a new phenomenology
and provides a novel research topic, both theoretically
and experimentally.
In fact, despite the simplicity of the model,
the trimer dynamics is affected by a strong
inner instability leading to the chaos onset.
This feature originates from the combination of
the model nonlinearity with the nonintegrable nature 
that distinguishes trimer dynamics within extended
regions of its phase space.
In respect of this, the solutions of the trimer dynamical
equations [displaying nonlinear periodic oscillations
(dimerlike orbits) with possible self-trapping effect]
recognized in Refs. \cite{NEM,FP3} must be viewed as
special subregimes in which dynamics is integrable.
Concerning the instability of trimer dynamics,
we wish to recall that the presence of homoclinic chaos
has been revealed for the asymmetric trimer in Ref. \cite{hg},
while, at the quantum level, the survival of breather
configurations has been investigated in Ref.~\cite{ff}.

In the present paper we perform
a systematic analysis of the trimer dynamics directed
to ascertain that the nonintegrability generates chaos
and that this dominates the interactions of three BECs
modeled in a standard semiclassical picture.
A rich scenario of dynamical behaviors emerges from our
analysis, which confirms the extremely structured
character of trimer dynamics and represents the natural
prosecution of the work of Ref. \cite{FP3} in which we 
focused our attention on the special dimerlike (integrable)
regime of trimer
and on the self-trapping effect. 

Indeed the analysis we perform appears to be 
topical in relation to the study of the dynamics of 
solitons~\cite{TROMB} and of vortices~\cite{CAPE} on 
one-dimesional chains of BECs as well as of experimental 
architectures obtained recently~\cite{INGU}. 
The observation of such chain
excitations, in fact, must take into account the possible
destructive action of inner instabilities whose influence
is clearly manifested in the trimer chain. We emphasize
the fact that the trimer chain is the simplest possible
situation in which interacting BECs turns out to be
governed by nonintegrable
equations. The dimer dynamics, in fact, is completely
integrable. 

The paper layout is the following.
In Sec. II we review the derivation of the space-mode
Hamiltonian for three coupled wells from the quantum field
theory of bosonic fluids in the dilute-gas approximation, 
%%%%%%%M
and present the semiclassical picture that describes 
coupled boson wells with macroscopic populations.
%
%and report on recent results (derived within this picture)
%on the integrable subregimes of trimer dynamics in which
%the self-trapping effect occurs.
%
In Sec. III we identify the set of fixed points of the trimer
Hamiltonian equations and show how such points are associated to
periodic solutions (representing collective modes)
owing to the dynamically conserved total boson
number $N$. Sec. IV is devoted to study the second variation 
(with $N =const$) of the energy function for such extremal 
configurations in order to make explicit their nature.
In Sec. V we perform a dimensional reduction of trimer dynamics 
by defining a new set of canonical variables whereby the constant 
of motion $N$ is incorporated in the dynamical equations. This 
paves the way to the implementation of the Poincar\'e sections' 
method.
The chaos onset in the trimer dynamics is investigated in Sec. VI.
First, in a qualitative way, by constructing the Poincar\'e 
sections for trajectories whose initial conditions are chosen 
in proximity of the fixed-point configurations identified in 
Sec. III. Then, quantitatively, by measuring the maximum Lyapunov 
exponent of such trajectories. 
Sec. VII contains concluding remarks and comments on future work.

\section{Trimer dynamics}

The model for a chain (or more complex structures) of
$M$ interacting bosonic wells can be derived from
the quantum field theory for boson fluids (with a nonlocal $\psi^4$
term) by implementing the M-(space)mode approximation~\cite{CIR}.
%
%The simplest situation involving a closed chain is $M = 3$.
%The semiclassical picture whereby
%the dynamical equations governing the trimer are obtained.
If the boson fluid is a diluite gas of $N$ interacting bosons
trapped in an external potential $V_{e}$ then its dynamics is
generated by the local boson-field Hamiltonian~\cite{PW,LEG,DGPS} 
\bq
{\hat {H}} =\!
\int
%{\cal s}
\! d{\bf r} \, \hat\psi^{^{\dagger}}({\bf r})
\left [ V_{e}
-\frac{_{\hbar^2 \nabla^2 }}{^{2m}} \!
+ \frac{_{U_0}}{^2} \hat\psi^{^{\dagger}}({\bf r})
\hat\psi({\bf r})
\right ]
\,\hat\psi({\bf r}) \non
\eq 
where $m$ is the boson mass,
$U_0=4 \pi \hbar^2 a/m$ takes into account
the two-body interaction strength, and $a$
is the $s-$wave scattering length. The field
$\psi({\bf r})$ [$\hat\psi^+({\bf r})$] is the Heisenberg
field operator that destructs [creates] bosons at position
${\bf r}$. The nonlinear term has been written in the usual
normal ordered form. In order to work out
the trimer Hamiltonian 
we state some assumptions:
First, $V_{e}$ is a three-well symmetric-shaped potential.
Second,
the lowest energy level of each well 
(within the approximation of $V_{e}$ in terms of a single-well parabolic
potential)
must be well separated from the higher energy levels~\cite{PW}.
Third, the binary particle interactions is not strong enough to
significatively
change the latter assumptions.

Some furher approximations are necessary
to make explicit the ground-state structure in $\hat H$.
Let ${\bf r}_i$ ($i=1,2,3$) be the locations of the minima of $V_e$
and let $V_j = V({\bf r}-{\bf r}_j)$ be the parabolic approximation
to the potential in the $j$th minimum, so that
$V_e({\bf r}) \simeq V({\bf r}-{\bf r}_j)$ when
${\bf r} \simeq {\bf r}_j$.
%
%$V({\bf r}) \simeq
%V_1({\bf r}-{\bf r}_1)+
%V_2({\bf r}-{\bf r}_2)+V_3({\bf r}-{\bf r}_3)$.
%
Also, let us introduce the eigenstates $u_j$
that represent the normalized single-boson ground-states
with energy $E_0$ of the local parabolic potential $V_j$. 
These states are only approximatively orthogonal because of
$\int d^3 {\bf r} \, {\bar u}_j u_k=\delta_{jk}+ R_{jk}$ but the residue
$R_{jk}$ is a quantity exponentially suppressed depending
on the overlap between $u_j$ and $u_k$.
The analysis is restricted to those potentials for which
$R_{jk}<<1$ thus making negligible such contributions.

The picture of the system thus resulting suggests that
the boson field operator can be expanded as
$\hat\psi({\bf r},t) = \Sigma_i
\bar{u}_i ({\bf r}) \hat a_i(t)$,
where $\hat a_i(t)$  [$\hat a^+_i(t)$]
is the annihilation [creation] boson operator (associated to
the space-mode state $u_i$) that satisfy the commutation relation
$[\hat a_i(t),\hat a^+_k(t)] = \delta_{ik}$.
By substituting this expression in the many-body hamiltonian
one can obtain, to the lowest order in the overlap between the
single-well modes, the quantum trimer Hamiltonian \cite{FPZ1,NEM}
$$
H_3= \Sigma_{i} [ U ( n_{i} -1)  - v] n_{i} -
\frac{{_T}}{{^2}}
\Sigma_{\langle i,j\rangle}
\left (a^{^{\dagger}}_{i} a_{j}+
a^{^{\dagger}}_{j} a_{i} \right ),
%\label{BHM}
$$
\noindent
where the site index $i,j=1,2,3$, $v =-E_0$, and the operators 
$n_{i}\doteq  a^{\dagger}_{i} a_{i}$ count the number of bosons 
at site $i$. Also, in $H_3$
$$
T= 2 \! \int \! d^3{\bf r} \, \bar{u}_j [V_j -V_{e} ]u_{j\pm1} 
\, , \,
U=\frac{_{U_0}}{^2} \int \! d^3{\bf r}|u_j|^4\, ,
$$
represent the (interwell) hopping amplitude and the
strength of the Coulomb on-site repulsion, 
respectively~\cite{MILB,PW}.
%whereas $v =-E_0$ the local parabolic potential. 
%and without loss in generality w can set equal to
%zero. 

%
%
The quantum dynamics involved by Hamiltonian $H_3$ can be 
cast in a classical form by representing the system quantum 
state through a trial state $| Z \rangle$ 
written in terms of Glauber's states $|z_i \rangle$ 
(defined by $a_i |z_i \rangle = z_i|z_i \rangle$). 
By implementing the
time-dependent variational principle (TDVP) 
and the procedure discussed in Refs. \cite{AP}, \cite{ZFG}
on $|Z \rangle= \Pi_i |z_i \rangle$,
one obtains
the effective
Hamiltonian~\cite{FPZ1}
$$
{\cal H}_3( Z, Z^*):= \langle Z| H_3 |Z \rangle \equiv
$$
\be
\Sigma^3_{j=1} \Bigl [ U |z_j|^4  -
v |z_j|^2 - \frac{_{T}}{^2}
\left (z^{*}_j z_{j+1} + 
%z^{*}_{j+1} z_j 
c.c.
\right ) \Bigr ] \,  ,
\label{HS3}
\ee
($j= 1 \equiv 4$ on the trimer chain) with the equations
\bq
%\cases{
 i\hbar {\dot z}_1 &=& (2U|z_1|^2 -v ) z_1
-\displaystyle \frac{_{T}}{^2} (z_2+z_3) \; , \non \\
%&${\-}$ \cr
 i\hbar {\dot z}_2 &=& (2U|z_2|^2  -v ) z_2 
-\displaystyle \frac{_{T}}{^2} (z_3+z_1) \; ,
\label{EM3}
\\
%&${\-}$ \cr
i\hbar {\dot z}_3 &=&  (2U|z_3|^2  -v ) z_3 
-\displaystyle \frac{_{T}}{^2} (z_1+z_2) \non
\;  .
\eq
Such equations for $z_j$ 
(notice that $z_j \equiv \langle Z| a_j|Z\rangle$
and $z^*_j = \langle Z| a^+_j |Z\rangle$) can be calculated from 
${\cal H}_3$ via the Poisson brackets 
$\{ z_k^* , z_j \}= i \delta_{k,j}/\hbar$ furnished by the TDVP 
method. Those for $z^*_j$ are easily obtained by complex conjugation.
Various aspects concerning the special dimeric subregime of
Eqs. (\ref{EM3}), in which trimer dynamics is integrable, have been
studied in Refs. \cite{NEM}, \cite{FP3}.
 
%%%%%%%%%%%%%%%%%%%%%%%%%%%%%%%%%%%%%%%%%%%%%%%%%%%%%%%%%%
\section{Fixed points and periodic orbits}

The distinctive features that characterize the dynamics of
a given Hamiltonian system can be recognized by exploring
the structure of its phase space.
The first step to do this is to locate the fixed points. 
The latter are derived in the present section based on
Eqs. (\ref{EM3}), whereas the nature of such points
is studied in the next section.

The fixed-point equations for the trimer are obtained by
setting ${\dot z}_j \equiv 0$ in Eqs. (\ref{EM3}). Since 
the trimer dynamics is constrained by the 
constant of motion $N= |z_1|^2 + |z_2|^2 + |z_3|^2$, one 
must incorporate explicitly the restriction to the phase-space
submanifold defined by $N = const$ through a Lagrange
multiplier $\chi$. This just requires that one
considers the variations of ${\cal H}_3 - \chi \, N$ in place of
that of ${\cal H}_3$.
The resulting equations are
\bq
0 = \left (2U |z_j|^2  -\mu + \frac{_T}{^2} \right ) z_j -
\frac{_T}{^2} Z \, ,
\label{FixP1}
\eq
where $j=1,2,3$, and $Z := z_1+ z_2+ z_3$,
$\mu : =\chi +v$ have been introduced .
Therefore, any fixed-point, namely any vector $(z_i) \equiv$
$(\eta_1, \eta_2, \eta_3)$ that satisfies Eqs. (\ref{FixP1}),
provides, at the same time, a dynamically active solution 
of Eqs. (\ref{EM3}) represented by $z_j (t) = \eta_j \exp[i\chi t/\hbar]$.
Despite its time dependence, such a {\em periodic orbit} represents
a (one-dimensional) extremal configuration of ${\cal H}_3$ on the
hypersurface $N = const$.
The solutions of Eqs. (\ref{FixP1}) can be grouped in various
classes the first of which is represented by the case
$$
Z= \Sigma_j\, z_j = 0 \, .
$$
The remaining cases are obtained from Eqs. (\ref{FixP3})-(\ref{FixP4}),
where $z_i$ can be replaced with the real quantities $x_i$ (see appendix 
\ref{fixed}). Such
cases are given by
$$
\begin{array}{c}
\!\!\!\!
x_1 \, = \, x_2 \, = \, x_3 \ne 0 \, , \cr
{\-} \cr
x_1 = - x_2 \ne x_3 = \, 0 \, , \cr
{\-} \cr
\!\!\!\! x_1 = x_2 \ne x_3 \ne \, 0 \, . \cr
\end{array}
$$
Such configurations are discussed below \cite{COM1}.
%%%%%%%%%%%%%%%%%%%%%%%%%%%%%%%%%%%%%%%%%%%%%%%%%%%%%%%%%%%%%%%%%%%%%%%%%%%%

\subsection{Ground-state configurations} 

\noindent
When $x_1 = x_2 = x_3$, Eqs. (\ref{FixP3}) and (\ref{FixP4}) 
are satisfied. Based on Eq. (\ref{MU}) 
combined with
the conserved quantity $N$, one finds that
$x_j = \pm \sqrt{N/3}$ which provides
\be
z_i =\sqrt{N/3}\, \exp(i\Phi) \, , \quad (i=1,2,3),
\label{MIN}
\ee
and $3 \mu = 2UN - 3T$.
The energy of such configurations (they are shown to represent the ground-state in Sec. III) is given by
\be
E_{gs}= \frac{_1}{^3} U N^2 -T N \, .
\label{Em}
\ee
The ground-state phase $\Phi$ is arbitrary
since it represents a symmetry of the model. Also,
the fact that $z_j = |z_j| \exp(i\phi_j)$ have the
same phase $\phi_j \equiv \Phi $ reproduces the
symmetry breaking phenomenon that distinguishes
the minimum energy state and, particularly, the
vanishing of the phase difference (between closed
points) in superfluid media.
Such two features, which are known to characterize
the superfluid ground-state of BH lattice model,
naturally extend to the trimer model of condensates
in our semiclassical picture.
As regards Eqs. (\ref{EM3}), they are solved by
$z_j(t) = \sqrt{N/3}\, \exp[ i( \Phi + \chi t/ \hbar)]$, 
with $\chi = \mu -v$, that describes the ground-state collective mode.
%
%%%%%%%%%%%%%%%%%%%%%%%%%%%%%%%%%%%%%%%%%%%%%%%%%%%%%%%%%%%%%%%%%%%%
\subsection{Vortexlike configurations}
\noindent
The situation in which $Z =0$ leads to a special configuration.
In this case Eqs. (\ref{FixP1}) reduce to
$0 = \left (2U |z_j|^2  -\mu + T/2 \right ) z_j$
implying, in turn, that
$$
|z_j|^2 =
\frac{_{2\mu - T}}{^{4U}} \; , \; \forall j \, , \quad \quad
\mu= \frac{_{4NU +3T}}{^6}
$$
where the value of $\mu$ is derived from
$\Sigma_i |z_i|^2 = N$. 
As a consequence of the independence of $|z_j|$ on the site index,
the condition $Z =0$ is realized only
if the phases of $z_j = |z_j| \exp (i \theta_j )$ are such that 
$\theta_j (k) = 2 \pi j k/3 + \Phi_0$ where $k = 1, 2$ and
$\Phi_0$ is an arbitrary phase. 
This configuration represents a particular case 
of the vortex state discussed for the Bose-Hubbard model on 
a $M$-well chain lattice ($M > 2$) in Refs. \cite{CAPE} and
\cite{FPZ1}. The energy associated to the vortex states
\be
z_j(k) = \sqrt{N/3} \exp [ i\theta_j(k)]
\label{VOR}
\ee
is given by
\be
E_{v} = \frac{_{U N^2}}{^{3}}
- T N \cos \bigl (\frac{_{2 \pi}}{^3} k \bigr ) 
= \frac{_{U N^2}}{^{3}} + \frac{_{TN}}{^{2}}
\, ,
\label{Ev}
\ee
while dynamics issued from Eq. (\ref{EM3}) is described by the 
solution $z_j (t)= z_j (k)\, \exp [it(\mu- v)/\hbar]$.
%%%%%%%%%%%%%%%%%%%%%%%%%%%%%%%%%%%%%%%%%%%%%%%%%%%%%%%%%%%%%%%%%

\subsection{Configurations with a single depleted well} 

These configurations are characterized by the presence of
a single depleted well (SDW).
Without losing generality, one can choose the second well,
so that one has $x_1,x_3\neq 0$, and $x_2=0$. This case can 
be faced based on Eqs. (\ref{FixP2}):
to satisfy $E_2({\bf x}) =0$ one must impose $x_3 = -x_1$,
which entails, in turn, that the equation $E_1({\bf x})=0$
is equivalent to the equation $E_3({\bf x})= 0$.
The latter, together with the
constraint $N = x_1^2 +x_3^2 $, implies
that
$x_1 = \pm \sqrt{{N}/{2}}= -x_3$,$x_2 = 0$,
which provides the fixed points
\be
\!\!
z_1= \sqrt{N/2}  e^{i \Phi}  , \,
z_2= 0  \, , \,
z_3= \sqrt{N/2} \, e^{i (\Phi +\pi)} ,
\label{SING}
\ee
with $
\mu \equiv 2Ux^2_1 + {T}/{2}$.
Their energy is
\be
E_{dw}= \frac{_1}{^2} U N^2 + \frac{_1}{^2} T N \,.
\label{Esdw}
\ee
Permutations of site indices $j$ of $z_j$ furnish other five
fixed points of the same type. In view of state (\ref{SING}),
it is worth noting that SDW configurations have the same
structure of the $\pi$-{\it states} occurring in the dimer
dynamics~\cite{SM1}, where the phases of each well keep 
a constant phase difference $\pi$ in the course of time 
evolution. Solutions $z_i(t) = z_i(0) \exp [it(\mu -v)/\hbar ]$
--with $z_i(0)$ given by Eqs. (\ref{SING})--
exhibit this property.

%%%%%%%%%%%%%%%%%%%%%%%%%%%%%%%%%%%%%%%%%%%%%%%%%%%%%%%%%%%%%%%%%%%%%%
\subsection{Dimerlike configurations} 

For such states the variables $x_j$ (recall that $x_j^2$
is the boson number of the $j$-th well) satisfy the condition 
$x_i = x_j \ne x_k$. Three cases are thus obtained through 
index permutation.
In general, since three fixed points turn out to be associated with 
each dimerlike configuration (namely to each choice $x_i= x_j \ne x_k$),
nine fixed points are finally found in the dimeric class.

To deal with an explicit case, we shall consider 
the fixed points of the case
$x_1 = x_2 \ne x_3$.
Owing to its complexity, their derivation is
described in appendix \ref{dimerFP}. 
Here we discuss the results obtained.
Fixed points depend on the parameter $\tau := T/NU$ and
are representable as
points $(x_1 , x_2, x_3) \in {\bf R}^3$ 
on a sphere due to the constraint $N = \Sigma_i x_i^2 $. 
Their expressions read
\bq
\begin{array}{c}
A_1 := \left (
a_1 , a_1, -(a_1/|a_1|) \sqrt{N-2 a_1^2}
\right ), \, \cr
{\-}\cr
A_2 := \left (
a_2 , a_2, -(a_2/|a_2|)\sqrt{N-2 a_2^2}
\right ), \, \cr
{\-}\cr
A_3 := \left (
a_3 , a_3, -(a_3/|a_3|)\sqrt{N-2 a_3^2}
\right ), \,
\end{array}
\label{VFP}
\eq
where
\be
a_3 =\pm
%\sqrt{ \frac{N p^2}{1 +2p^2}  }
\left [ \frac{N p^2}{1 +2p^2} \right ]^{\frac{1}{2}}
%[Np^2/(1 +2p^2)]^{1/2},
\, , \,\,
a_\nu= \pm
%\sqrt{ \frac{N}{2 +q^2_\nu} }
\left [ \frac{N}{2 +q^2_\nu} \right ]^{\frac{1}{2}} ,
%[N/(2 +q^2_\nu)]^{1/2},
\label{PARF} 
\ee
and $\nu=1, 2$. Parameters $q_\nu$ and $p$ are defined as implict 
functions of parameter $\tau$ through systems (\ref{EQ}) and 
(\ref{EP}), respectively. In particular, the cubic equation 
\be
\tau (2+q^2)(2+q) + 4q\, (1+q) =0 \, ,
\label{CQ}
\ee
derived from Eq. (\ref{EQ}), with  $q \in [-1,0]$ furnishes
the $\tau$ functions $q_1(\tau)$, $q_2(\tau)$ corresponding to
the real roots of Eq. (\ref{CQ}). Notice that
$-1 \le q_1(\tau) < q_2(\tau) \le 0$ for $\tau < \tau^*$, 
whereas, for $\tau > \tau^*$, there are no solutions. The value $\tau^*$, 
where $q_1(\tau^*) = q_2(\tau^*)$ (and $A_1 \equiv A_2$), is calculated 
in appendix \ref{dimerFP}.
Instead, system (\ref{EP}), with $p \in [-1/2, 0]$,
always exhibits a single solution $p(\tau)$ that is
carried out from the equation
\be
\tau (1+2 p^2)(1+ 2p) + 4p\, (1+ p) =0 \, .
\label{CP}
\ee
The dependence on physical parameters $T$, $U$, $N$ of $A_1 (q_1)$,
$A_2 (q_2)$, $A_3 (p)$ by means of parameter $\tau$ is thus established. 
%%%%%%%%%%%%%%%%%%%%%%%%%%%%%%%%%%%%%%%%%%

Points $A_i$ found in this way generate, by varying $\tau$, three 
curves on the sphere with $N= const$ [in view of the sign $\pm$
in Eqs. (\ref{PARF}) they actually are six].
These become eighteen when considering the fixed points
generated 
by index permutations. This process is described in
appendix \ref{dimerFP}, where the actual number of dimerlike fixed 
points
is shown to reduce to twelve.
Such curves [parametrized by $\tau$ via $p(\tau)$ and $q_{\nu}(\tau)$] 
can be proven to never intersect with one another except for the
special case $\tau = \tau^*$ where $A_1 \equiv A_2$. This 
{\it coalescence}
effect is discussed below.

If the values of $q_\nu$ and $p$ for some given $\tau \le \tau^*$
are carried out explicitly, the energy for $A_i$ 
\be
E_{d} = U[N^2 + 6 a^4_{i} - 4 N a^2_{i} ]-
T [a^2_{i} -2 a_{i} \sqrt{N-2 a^2_{i}} ] \, ,
\label{Edim}
%Em-Ev-Esdw-Edim
\ee
($i =1,2,3$) is obtained via formulas (\ref{PARF}). 

We conclude illustrating the physical situations
that correspond to configurations $A_1$, $A_2$ and $A_3$ 
when $\tau$ changes. Also, we compare them to the 
{\it pure-dimer} scenario \cite{SM1, FPZ1}.
Let us start with $\tau \to 0$. One has $q_1=-1$ and  $q_2= p =0$
that entail
\bq
\begin{array}{c}
A_1 (-1) := \pm \left (\sqrt{N/3}, \sqrt{N/3}, 
-\sqrt{N/3} \right ), \cr
{\-}\cr
A_2 (0) := \left ( \sqrt{N/2} , \sqrt{N/2}, \, 0 \right ), \cr
{\-}\cr
A_3 (0) := \left (0,\, 0, \, \sqrt{N} \right ) \, ,
\end{array}
\label{CONF1}
\eq
respectively.
By increasing $\tau$, $A_1 (q_1)$ and $A_2 (q_2)$
get closer and closer ($A_1 = A_2$ for $\tau \to \tau_* $). 
When $\tau > \tau_* $ only the fixed point
$A_3 (p)$ survives.
In particular, 
$\tau \to \infty$ implies $p = -1/2$, so that
[from the third equation of Eqs. (\ref{VFP})] one obtains 
\be
\!
A_3 (-\frac{_1}{^2}) = \pm \left (\sqrt{N/6}, \sqrt{N/6}, 
-\sqrt{2N/3} \right ) .
\label{CONF2}
\ee
Since $A_3(p)$ [as well as $B_3(p)$, 
$C_3(p)$, obtained via index permutation (see appandix 
\ref{dimerFP})] is shown to be a maximum in Sec. IV, indeed
$A_3(p)$ appears to be comparable with the $\tau$-dependent maximum 
of the (pure) dimer model~\cite{FPZ1}, where a unique well ends up 
to be filled when $\tau \to 0$.
Instead, when $\tau$ is increased, no merging of $A_3(p)$
with other maxima [e.g., $B_3(p)$, $C_3(p)$]
happens as that observed in the (pure) dimer model.
In this model, in fact, a (macroscopic) {\it coalescence}
effect takes place [see Ref. \cite{SM1, FPZ1}] since two maxima
merge in a unique one when $\tau > 1$ [the opposite effect
(bifurcation) 
occurs for $\tau < 1$]. As shown in Appendix A, such effects 
involving maxima pairs do not distinguish trimer dynamics.

A different macroscopic effect happens, however,in the 
trimer phase space. This is caused by the merging of distinct saddle 
points (e. g., $A_1$, $A_2$) discussed above, that disapper for 
$\tau > \tau^*$. 
Since chaotic behavior develops around saddle points, their coalescence 
should entail an evident local regularization of dynamical behavior.

As a final remark, we wish to observe that
state $A_1(-1)$ exhibits the same per-well boson distribution
of the ground-state. Such two states differ uniquely owing to the phase
of the third well. Similarly, state $A_1(0)$ and the $\pi$-like 
state $(\sqrt{N/2}, -\sqrt{N/2} , 0)$ just differ owing to the opposite 
phase of the
second well.
Despite their identical boson distribution, such situations will 
exhibit very different behavior thus confirming the profound influence 
of the interwell phase differences in distinguishing dynamical states.

\noindent
%\hrulefill

%%%%%%%%%%%%%%%%%%%%%%%%%%%%%%%%%%%%%%%%%%%%%%%%%%%%%%%
\section{Character of fixed points}

In this section we consider the stability character 
of fixed points just identified.
Such a character is recognized by studying the second variation 
of the energy function on the hyper-surface defined by $N=const$. 
Explicitly, one should analyze the signature of the
quadratic form associated to the Hessian of ${\cal H}_3$
(with $N =const$) in each fixed point.
Being this process 
rather technical, we mainly develop it in Appendix \ref{charact}. 
Below, after showing how the separation of ${\cal H}_3$ in two independent 
(local) subhamiltonians simplifies remarkably the stability analysis,
we summarize the results obtained.

To simplify ${\cal H}_3$, it is advantageous to introduce 
the local
variables $\xi_j =z_j - v_j$, where $v_j$ are the 
coordinates
of some given fixed point.
Then, neglecting third and fourth order terms, ${\cal H}_3$ 
takes the form
$$
{\cal H}_3 \, = {\cal H}_3(v) +
\Sigma^3_{i=1} \Bigl (2U |v_j|^2 -\mu - T \Bigr ) |\xi_i|^2 
$$
\be
+\Sigma^3_{j=1}  U (v^*_j \xi_j + v_j \xi^*_j)^2 +\frac{_T}{^4}
\Sigma_{i\neq k} |\xi_i - \xi_k|^2 ,
\ee
%%%%%%%%%%%%%%%%%%%%%%%%%%%%%%%%%%%%
%\bq &{\cal H}_3& \, = {\cal H}_3(v) +
%\Sigma^3_{j=1} \Bigl [ (2U |v_j|^2 -\mu )|\xi_j|^2 
%- T |\xi_i|^2 \Bigr ] \non \\ &&\Sigma^3_{j=1}  U (v^*_j \xi_j + v_j \xi^*_j)^2
%+ \displaystyle \frac{_T}{^4} \Sigma_{i\neq k} |\xi_i - \xi_k|^2 \, ,
%\eq
which undergoes the further semplification
$v_i =$ $ v^*_i \to x_i$ when one recalls that the phase factor of
$v_i \in {\bf C}$ is a constant that can be absorbed by $\xi_j$.
This fact allows us to separate
${\cal H}_3$ in a $q$-dependent part and a $p$-dependent part,
with $\xi_j := q_j + i p_j$. By making explicit 
the latter definition in ${\cal H}_3$, we find 
\be
{\cal H}_3 = {\cal H}_0 + h(q\, ; 6U, T, \mu) +
h(p\, ; 2U, T, \mu) 
\label{H3hh} 
\, ,
\ee
where ${\cal H}_0 :={\cal H}_3(v)$, and
\be
h(q\, ; 6U, T, \mu) :=
%
%%%\Sigma^3_{j=1} \left [ (6U x_j^2 -\mu )\, q_j^2 \right ]
%%%- T \left ( q_1 q_2 + q_2 q_3 +  q_3 q_1 \right ) \, ,
\sum_{ij} (M_q )_{ij} \, q_i q_j \, ,
\label{hq}
\ee
\be
h(p\, ; 2U, T, \mu) :=
%%%\Sigma^3_{j=1} \left [ (2U x_j^2 -\mu )\, p_j^2 \right ] 
%%%- T \left ( p_1 p_2 + p_2 p_3 +  p_3 p_1 \right ) \, ,
\sum_{ij} (M_p )_{ij} \,  p_i p_j \, .
\label{hp}
\ee
Dynamical matrices $M_q$ and $M_p$ are defined as
\be
M_q=
-\frac{T}{2} \left [ \begin{array}{ccc}
\Delta_1 &1 &1 \\
1 &\Delta_2 &1 \\
1 &1 &\Delta_3 \\
\end{array} \right]
\; ,
\ee
\be
M_p=
-\frac{T}{2} \left [ \begin{array}{ccc}
\delta_1 &1 &1 \\
1 &\delta_2 &1 \\
1 &1 &\delta_3 \\
\end{array} \right]
\, ,
\ee
with $\Delta_j := 2(\mu -6Ux_j^2)/T$ and
$\delta_j := 2(\mu- 2Ux_j^2)/T$.
Hence the diagonalization of the (local) quadratic form 
associated to ${\cal H}_3$
can be performed in a separate way on (\ref{hq}) and (\ref{hp}).
Further simplifications come from the fact that
matrix $M_p$ 
is proven to always have a vanishing eigenvalue (see Appendix 
\ref{zeroeig}), while, due to the conserved quantity 
$N = \Sigma_i |z_i|^2 = const$, the induced local constraint 
$q_1 x_1 + q_2 x_2 +q_3 x_3 = 0$ makes $h(q\, ; 2U, T, \mu)$
dependent only on two variables $q_i$ (see Appendix 
\ref{charact}).

%%%%%%%%%%%%%%%%%%%%%%%%%%%%%%%%%%%%%%
The stability character for ground-states,
vortex states, SDW states, and dimerlike states
is studied explicitly in Appendix \ref{charact}. The
calculation of the ${\cal H}_3$ second variation and,
when this is necessary, of Hessian eigenvalues provide
the following scenario.

\medskip
\noindent
- States with $x_1 =x_2 =x_3$ are energy minima;

\medskip
\noindent
- Vortex configurations are saddle points;

\medskip
\noindent
- SDW configurations are saddle points;

\medskip
\noindent
- Dimeric states exhibit two saddle points and one
{\-} $\,$ maximum point for $0<\tau<\tau^*$.

\medskip
\noindent
In the last case the merging of the two saddle points
to form a regular point is enacted for $\tau \to \tau^* $.
Thus, for $\tau^* < \tau$, a single maximum survives.
%
%
%%%%%%%%%%%%%%%%%%%%%%%%%%%%%%%%%%%%%%%%%%%%%%%%%%%%%%%%%%%%%%
%
\section{Mapping of dynamics on the reduced phase space}

%The remaining part of present work is devoted to develop
We develop a both qualitative and quantitative 
analysis of the chaotic behavior of the trimer
based on the Poincar\'e section method.
After performing a qualitative study of the periodic-orbit 
instability, we effect a quantitative analysis
by measuring the very chaos' indicator, namely the Lyapunov
exponent of every single orbit choosen near a period one.
To this end, we introduce a more suitable
coordinate 
system involving a symplectic reduction of dynamics.

Hamiltonian (\ref{HS3}) depends on three complex
degrees of freedom
and commutes with the boson number $N$. This fact as well as
the structure of coupling term in ${\cal H}_3$ permit us
to reduce from six to four the number of (real) canonical 
coordinates. 
The numerical integration of Eqs. (\ref{EM3}) thus furnishes 
a system picture consisting of a trajectory in a 
four-dimensional ($4d$) {\it reduced} phase-space 
${\cal P} \in {\bf C}^3$. 
In ${\cal P}$, a Poincar\'e section (PS) is
the figure made by the points where a trajectory cut a
$2d$ reference plane.
The new set of canonical coordinates used to construct the PS
embodies explicitly the conserved quantity $\Sigma_{i=1}^{3} 
|z_i|^2 = N $. Complex coordinates 
$z_i = \sqrt{n_i} \exp (i \theta_i)$
are replaced by
\begin{equation}
%\left\{
\begin{array}{ll}
\phi_1 = \theta_2 - \theta_1 \ , 
&
\xi_1 = 
(n_2 +n_3 -n_1)/N \\
%1 - 2 n_1 /N \ , \\ 
\\
\phi_2 = \theta_3 - \theta_2 \ , 
&
\xi_2 = 
(n_3- n_1 -n_2)/N \\
%2 n_3/N -1 \ , \\
\\
\psi = (\theta_3 + \theta_1)/2 \ , &
N = n_1 + n_2 + n_3 \  , 
\end{array}
%\right.
\label{chart}
\end{equation}
which obey the canonical Poisson brackets
\begin{equation}
%\left\{
\begin{array}{ll}
\{ \phi_i, \xi_j \} = -2 \delta_{ij}/N \ , &
\{ \phi_i, N \} = 0 \ , \\ \\
\{ \psi, \xi_j \} = 0 \ , &
\{ \psi, N \} = - 1 \ . 
\end{array}
%\right.
\end{equation}
%
%-->COM2
%It is important to stress the fact that the coordinate change 
%expressed by definitions (\ref{chart}) is well defined only for 
%$z_1,z_2,z_3\neq 0$. Therefore, situations where such particular 
%points are involved --fixed points of the $\pi$-state case provide 
%an example of this pathology-- must be dealt with some caution.
%
With such new variables~\cite{COM2}
%Hamiltonian (\ref{HS3}) assumes the form 
${\cal H}_3$ becomes
$$
%\be
%\begin{array}{q}
{\cal E}= \displaystyle
\frac{_2}{^{U N^2 }} {\cal H}(\xi_1,\xi_2,\phi_1,\phi_2)
 =
$$
$$
\xi_1^2 + 
\xi_2^2 - \xi_1 \xi_2 - \xi_1 + \xi_2 + {\cal E}_0  -
%\\   
%\quad \quad \quad \quad \quad \quad \quad \quad \quad 
$$
$$
\tau \sqrt{(1-\xi_1)(\xi_1-\xi_2) (1+\xi_2)} \times
$$
\be
\left[
% \displaystyle
\frac{\cos \phi_1 }{\sqrt{(1+\xi_2)}} +
% \displaystyle
\frac{ \cos \phi_2 }{\sqrt{(1-\xi_1)}} +
% \displaystyle
\frac{ \cos \phi_{12} }{\sqrt{(\xi_1-\xi_2)}} 
\right]
\label{H4} 
%\end{array}
\ee
in which $\phi_{12} := \phi_1+\phi_2$, ${\cal E}_0 = 
1 -2\mu/(UN)$ 
%
%that incorporates explicitly the conserved quantity $N$.
%
(the associated Hamiltonian equations are contained 
in Appendix \ref{fprc}).
In terms of coordinates $(\xi_1,\xi_2,\phi_1,\phi_2)$
the ground-state configuration (\ref{MIN}) and 
vortexlike fixed points (\ref{VOR}) correspond to
$$
\left ( \frac{_1}{^3},-\frac{_1}{^3},\, 0,\, 0 \right )\, , \quad
\left( \frac{_1}{^3},-\frac{_1}{^3}, \frac{_2}{^3} \pi k,
\frac{_2}{^3} \pi k \right ) \, ,
$$
($k=1,2$), respectively.
Dimeric fixed points 
[consider, e. g., $A_1$ in Eq. (\ref{VFP})] are given by 
$$
(\xi_1, 2{\xi_1}-1 ,0,\pi)\, ,
$$
where ${\xi_1}=1- 2 a^2_1 (\tau) /N$ and
$a_1 (\tau)$ is defined by Eqs. (\ref{PARF}), whereas SDW states
can be expressed as $(0, 0, \varphi, \pi- \varphi )$ for $n_2 =0$.
%Also, singular points, in which transformations (\ref{chart}) 
%are illdefined, are discussed in Appendix \ref{fprc}.

Operationally, the motion equations (\ref{EM3}) are numerically 
integrated by using a first-order bilateral sympletic scheme; 
the algorithm precision is checked by monitoring the conserved 
quantities, that is, the system energy and total number of bosons.
Trajectories can be traced in the phase space $\cal P$ in terms of
$\xi_1$, $\xi_2$, $\phi_1$, and $\phi_2$. For 
any given value of the reduced energy, Hamiltonian (\ref{H4}) defines
a 3d hypersurface in $\cal P$. The 2d surface used to construct 
a PS then is obtained by firming the value of $\xi_2$ to a constant. 
Hence, PS is made by coordinates $(\xi_1,\phi_1)$ 
of the points' set in which trajectories cut
the selected 2d surface.
%
%In addition to the PS of hypersurfaces close 
%to the fixed points, the next section contains the results of the 
%quantitative study of the chaotic behavior achieved by measuring 
%the Lyapunov exponent of every single trajectory.

%%%%%%%%%%%%%%%%%%%%%%%%%%%%%%%%%%%%%%%%%%%%%%%%%%%%%%%%%%%%%%%%%%
\subsection{Discussion of numerical results}

We present the results of the numerical analysis aimed 
at investigating the phenomenology of the trimer dynamics
in proximity of fixed points -these identify with the periodic
orbits that stationarize ${\cal H}_3$ with $N = const$-
calculated in Sec. III. 
The Hamiltonian parameters and the total boson number
chosen for the numerical simulations are
$$
U = \, T =\, 1 \, ,\quad v =0 \,, \quad N = \, 10\, ,
$$
respectively. Simulations have been carried out by using an 
integration time step of order $1 \times 10^{-4}$, while the 
total number of time steps employed in constructing each orbit
is of order $2^{28}$.
For each periodic orbit a PS has been selected by setting
$\xi_2=const$ and $d{\xi_2}/dt >0$; on this section we have
considered samples of about 100 initial conditions (IC).
Also, the
maximum Lyapunov exponent (MLE) has been measured for each
trajectorie of the samples associated to the extremal periodic
orbits.
\medskip

{\it Ground-state}.
The PS of this case is fixed by the condition $\xi_2= -1/3$.
About 100 initial conditions have been choosen in the
$\xi_1\! -\! \phi_1$ plane, placed close to the point
$(\xi_1, \phi_1) = (1/3, 0)$.
The energy of the corresponding trajectories is
$E\simeq 0.47 (UN^2/2)$. Fig. \ref{secgs} shows that
all the trajectories are regular in the phase space $\cal P$.
Their regular character appears to be consistent with
the periodic character shown by the
oscillations of populations $n_i=|z_i|^2$
(see Fig. \ref{popgs}) of each condensate.
\medskip
%%%%%%%%%%%%%%%%%%%%%%%%%%%%%%%%%%%%%%%%%%%%%%%%%%%%%%%%%%%%%

{\it Vortexlike initial conditions}.
The choice $\xi_2= -0.3$, $E \simeq 0.77 (UN^2/2)$
distinguishes the PS of the present case. The initial
conditions for the trajectories have been chosen close to
the fixed points $(\xi_1, \phi_1) =$ $(1/3, 2k \pi/3)$,
$k =1,2$ (vortex state). Since no difference distinguishes
the PS with $k=1$ and that with $k=2$, we restrict
our attention to $k=1$.
Fig. \ref{secvrtx} shows the presence of
both regular and chaotic trajectories [Fig. \ref{secregchavrtx}
supplies two examples, one for each orbit type].
%
%[in particular, Fig. \ref{secregvrtx} (Fig. \ref{secchavrtx})
%represents a regular (chaotic) orbit].
%
It shows as well that the PS points related to chaotic
trajectories are distributed in a region well separated
from that occupied by points generating regular orbits.
%
%Figs. \ref{secregvrtx} and \ref{secchavrtx}
%are illustrated in Fig. \ref{lypregchavrtx}.
%
In particular, the PS displayed in Fig. \ref{secvrtx}
[together with other PS involving slightly different
$\xi_2$ ($\simeq -1/3$)] suggests that the vortex-state
fixed point is basically surrounded by chaotic orbits.
%Thus, the corresponding periodic orbit seems to be strongly unstable.
%Apparently, the periodic character of vortex state does
%not survive, neither for short times, within its neighbourhood.
%
In Fig. \ref{secvrtx}, to reach the nearest regular orbits 
starting from $\xi_1=1/3$, $\phi_1 = -2\pi/3 $,
a finite variation of both $\xi_1$ and $\phi_1$ is necessary.

%Consider, e. g., the highest orbit of the regular ones
%in Fig. \ref{secvrtx}.
%
When this change is carried out the trimer-population
oscillations change in a significant way.
The time evolution of condensate populations $n_i$ ($i=1,2,3$)
related to the nonchaotic orbit of Fig. \ref{secregchavrtx}
is illustrated in Fig. \ref{popregvrtx} and confirms its regular
character.
By considering IC closer and closer to the vortex state position
such a character is progressively lost.
This is shown in Fig. \ref{popchavrtx} that plots
populations $n_i$, as a function of time, for the
chaotic trajectory of Fig. \ref{secregchavrtx}.
%(the IC of the latter are close to the vortex state position).
%

The regular orbit of Fig. \ref{secregchavrtx} involves
an evident {\it self-trapping} effect provided their IC
are enough far from $(\xi_1, \phi_1 )= (1/3 , -2\pi/3)$.
This is clearly
manifested in Fig. \ref{popregvrtx} where $n_i(t)$'s
oscillate in such a way that $n_2(t) < n_1(s)$, $n_3(s)$,
$\forall t$, $\forall s$: a {\it stable gap}, 
in fact, separates $n_2$ from $n_1$, $n_3$.
On the contrary, no stable gap is involved, in general,
by chaotic orbits (see Fig. \ref{popchavrtx}) that
develop large oscillations on the whole range of $n_i$.
The fact that, in average, $n_i \simeq 1/3$ is the only
feature inherited by the vortex state.
\medskip
%%%%%%%%%%%%%%%%%%%%%%%%%%%%%%%%%%%%%%%%%%%%%%%%%%%%%%%%%%%%%%%
%

{\it SDW-like intial conditions}.
In this case the choice $\xi_2=-0.002$ and 
$E\simeq 1.09 (UN^2/2)$ fixes the PS
that is represented in Fig. \ref{secoew}.
The zoom of the section reveals regular trajectories
in $\cal P$ placed near SDW fixed points
the latter being characterized by $\xi_2 = 0 =\xi_1$,
$\phi_1 = \varphi$, $\phi_2 =\pi- \varphi$, where we
have set ($\varphi$ can be choosen arbitrarily)
$\varphi = 0.57 \, \pi$.
Fig. \ref{secchaoew} describes the PS of a chaotic
orbit chosen among those of Fig. \ref{secoew}:
The PS points are distributed in two, well separated,
basins in a quite evident way. The interpretation of
%Fig. \ref{secchaoew}
such an effect is the following:
after recalling that setting $\xi_2 \simeq 0$ implies that
$n_3 \simeq 1/2$, one deduces that, concerning the points of
the PS, the values allowed for $n_1$
are either $n_1 \simeq 1/2 $ or $n_1 \simeq 0 $,
which involves either $n_2 \simeq 0 $ or $n_2 \simeq 1/2 $,
respectively.

One therefore recognizes the presence of an {\it inversion population}
phenomenon beetwen $n_1$ and $n_2$. Interestingly, no intermediate
values seems to be permitted.
The corresponding scenario is given, on a shorter time interval,
in Fig. \ref {popchaoew} where the nonperiodic oscillations
of populations $n_1(t)$, $n_2(t)$, and $n_3(t)$ are
compared and a population inversion effect
involving $n_1$, $n_2$ gets going.
The population oscillations referred to a regular orbit
of those contained in the zoom of Fig. \ref{secoew} are shown in
%%%%%%\ref{secregoew} and
Fig. \ref{popregoew}.
%in Fig. \ref{lypregchaoew}.
\medskip
%%%%%%%%%%%%%%%%%%%%%%%%%%%%%%%%%%%%%%%%%%%%%%%%%%%%%%%%%%%%%%%%%

{\it Dimerlike initial conditions}. As proven in Secs. III
and IV, the fixed points of this case consist of two saddle
points and a maximum. 
The conditions $\xi_2=-0.295$, $E\simeq 0.73 (UN^2/2)$
and $\xi_2=-0.005$, $E \simeq 0.91 (UN^2/2)$
firm the PS associated to the first saddle
(see Fig. \ref{secsella1}) with coordinates
$\xi_1 = -\xi_2 =0.295$, $\phi_1 = -\phi_2 = \pi$,
and to the second saddle
(see Fig. \ref{secsella2})
with coordinates
$\xi_1 = -\xi_2 = 0.005$, $\phi_1 = -\phi_2 = \pi$,
respectively. In both the cases,
the PS's exhibit both regular and chaotic trajectories.
Concerning Fig. \ref{secsella1} (first saddle
point), the coexistence of such regimes is confirmed in
Fig. \ref{secregsella1},
%and \ref{secchasella1}
% and \ref{lypregchasella1}
where the PS of a regular trajectory is compared
with the PS of a chaotic one. Figs. \ref{secregsella2}
%and \ref{secchasella2}
%\ref{lypregchasella2}
shows analogous quantities referred to the second saddle point.
It is worth noting that in the latter case the neighbourhood
of the saddle point is characterized IC issuing regular orbits, 
whereas the first saddle is surrounded by IC generating chaotic
motions.
%%%%%%%%%%%%%

In Fig. \ref{secsella2}, regular orbits reside either on the
bottom or on the top of the figure, but dynamics never connects
the top orbits with the bottom ones. This feature is confirmed
by the time behavior (see Fig. \ref{popregsella2}) of $n_i (t)$
related to the regular trajectory of Fig. \ref{secregsella2}.
In Fig. \ref{popregsella2}, the presence of the gap beetwen
$n_2\, (\simeq 0.5)$ and $n_1 $, $n_3$ $(\simeq 5)$ indicates
a macroscopic {\it self-trapping} effect. The latter differs from
the dimeric self-trapping reviewed in Sec. II-A, where
$z_1= z_3 \to n_1 = n_3$, in that $n_1$, $n_3$ develop
{\it independent} oscillations.
The scenario just described no longer holds for the chaotic
orbits: Fig. \ref{popchasella2} shows an {\it intermittent}
effect of population inversion beetwen $n_1$ and $n_2$.
This reflects the fact the points of the chaotic orbit of
Fig. \ref{secregsella2} are distributed intermittently 
both in the higher and in the lower part of the PS.

The trajectories near the maximum
(we consider the case $\xi_1 = 0.999$, $\xi_2 =0.998$,
$\phi_1 = -\phi_2 = \pi$,) appear to be regular, as
one can deduce from Fig. \ref{secMax} that shows PS close to 
a maxium fixed point with $\xi_2=0.8$ and $E\simeq 1.7 (UN^2/2)$.
In this case, populations $n_1(t)$, $n_2(t)$, and $n_3(t)$ display
in Fig. \ref{popMax} a periodic effect of
{\it population-inversion} involving $n_1$, $n_2$, whereas
the fact that $n_1$, $n_2$ $<< n_3$ entails an evident
self-trapping phenomenon.

The averages of the MLEs calculated for the chaotic orbits are
%%%secvrtx/secoew/secsella1/secsella1
0.74 (Fig. \ref{secvrtx}),
0.42 (Fig. \ref{secoew}),
0.29 (Fig. \ref{secsella1}), and
0.96 (Fig. \ref{secsella2}).
The evaluation of the MLE for the regular trajectories described
in the previous examples (as well as for those of Fig. \ref{secgs}
and Fig. \ref{secMax}) exhibit the expected decreasing behavior
thus suggesting that
%(see Fig. \ref{lyapMax})
a weak stochasticity occurrs on such trajectories. 

%%%%%%%%%%%%%%%%%%%%%%%%%%%%%%%%%%%%%%%%%%%%%%%%%%%%%%%%%%%%%%
\section{Conclusions}

In this paper we have focused our attention on the
structure of the phase space of 'classic' trimer,
that is 
the mean-field form of the model describing
three interacting 
BECs. Our analysis puts in light, on the one hand,
the 
remarkable complexity that characterizes the trimer
dynamics (by comparison with the integrable dynamics of
the dimer system), on the other hand, the phenomena that
are expected to characterize the $M$-well chain of
interacting condensates. In view of the recent experimental
results, the phenomenology of this system seems
more and more 
viable to experimental observations.
Trimer dynamics has been investigated
within a semiclassical Hamiltonian picture,
reviewed in Sec. II and developed in previous papers,
based on a coherent-state representation of the
trimer quantum state. 

The identification of the set of fixed points of
trimer Hamiltonian equations and the fact that such
points are associated to periodic solutions (collective
modes) of several types represents the initial, central
result of our paper. The presence of the constraint $N= const$
entails that the states that stationarize the Hamiltonian
are not isolated points but periodic orbits (one-dimensional
manifolds).
The solutions thus found enlarge the set of exact 
solutions~\cite{NEM,FP3}
pertaining the dimerlike integrable 
subregimes of trimer
and exhibiting a parameter-dependent 
{\it self-trapping} effect.

Based on the second variation (with $N = const$) of
the energy function around its fixed points,
the character of the latter has been recognized in Sec. IV
and appendix \ref{charact} thereby revealing the presence 
of several saddle points
and maxima, in addition to the expected ground-state.
Numerical simulations and the PS method have furnished 
a wide scenario of trimer dynamical
behaviors whose 
possible {\it chaotic} character has been tested
by measuring the maximum Lyapunov exponent. 
We summarize the results of our dynamical simulations.

\noindent
{\bf i}) The orbits that have ICs close to ground-state exhibit
a regular behavior with periodic oscillations of populations
$n_i$. 

\noindent
{\bf ii}) On the contrary, orbits with vortexlike IC (namely based
at points close to vortex fixed points) are, in general, chaotic.
Regular orbits, however, are found at sufficient distance from
vortex fixed points.
For such orbits a stable gap separates the oscillations of
$n_i$ from those of $n_j$, $n_k$ ($j,k \ne i$) thus generating
{\it self-trapping}; $n_j$, $n_k$ show independent oscillations
(in Sec. VI configurations with $i=2$, $j=1$, $k=3$ has been
considered). The gap disappears for chaotic orbits.
%Equivalent behaviors are obtained 
%by permutation of site indices.

\noindent
{\bf iii})
Orbits with SDW-like ICs also display both regular and
chaotic behaviors but their ICs are not separated spatially.
The regular orbits we have considered keep memory
the IC since one of the three wells remains almost empty
(pure dimer) while the other two undergo regular oscillations.
Such states identify essentially with $\pi$-{\it like} 
{\it states} and
manifest a stable character.
An example of {\it chaos emergence}, 
which starts with a macroscopic
{\it population inversion}  
entailing the filling of the (initially)
depleted well, has been detected by assuming
various SDW-like ICs involving a chaotic orbits.

\noindent
{\bf iv}) Regular orbits generated by dimerlike ICs
(related to the
second saddle point) exhibit periodic 
oscillations of
$n_i$'s with an evident {\it self-trapping}.
Chaotic orbits, instead, give rise to oscillations displaying
self-trapping on short time intervals and {\it intermittent}
population-inversion effects. ICs close to the maxima
further generate regular orbits with self-trapping.
The corresponding states display the presence of a
unique almost filled well~\cite{ff}.

The scenario just depicted supplies a rich account of
properties, behaviors and possible observable effects issuing
from trimer dynamics and suggests promising future developments.
An effect that might have a macroscopic character is related 
to the expected chaos suppression (see Sec. IIID) 
caused by the $A_1$-$A_2$ coalescence for $\tau = \tau_*$. 
The basic configurations (ground-state, vortex states,
SDW states, and dimerlike states involving both saddle 
and maximum
points) recognized in the present work,
and the complexity of dynamical regimes,
both chaotic and regular, that develop in their neighbourhoods
deserve further investigations in two directions at least.
First, classically, one should carry out
a systematic study (requiring huge computational resources) of
long-time behavior of dynamical states of interest to disclose
further macroscopic effects. For the same reason, a larger number
of ICs (together with the trajectories thus issued)
should be considered near fixed-point configurations.

We point out the fact that predictions on the dynamics
of phases $\phi_i$ should be important in relation to
phase-interference experiments \cite{LATT}.
This aspect, which has not been deepened in the present work,
requires a separate analysis and further numerical study
directed to detect phenomena exhibiting phase coherence
and their stability in proximity of states
endowed with ordered phase configurations
such as vortex states
($\phi_i = 2\pi k/3 $, $k=1, 2$),
SDW states ($\phi_i - \phi_j = \pi$, $i\ne j$,
$|z_k|=0$ with $i, j\ne k$),
and dimer configurations
($\phi_i = \phi_j$, $z_k \ne z_i= z_j$ with $i, j\ne k$).

Second, in view of the possibility of realizing
systems with small per-well populations, the pure
quantum approach to trimer dynamics (along the same lines
of previous work directed to study the spectral properties of
dimer) seems to be quite natural. The study of quantum
trimer might put in evidence unexpected effects caused
by the competition of chaotic (classical) behavior and
integrable (quantum) behavior on the borderline of
appropriate mesoscopic regimes where the transition
from quantum to classical dynamics takes place.
%
%
%%%%%%%%%%%%%%%%%%%%%%%%%%%%%%%%%%%%%%%%%%%%%%%%%%%%%%%%%%%%%%%%
\acknowledgments
We are indebted to M. Pettini for useful observations
and to P. Buonsante for his comments.
The work of V.P. has been supported by I.N.F.M. (Italy) and
M.U.R.S.T. (within the Project "Theory of coherent fluids").
R.F. gratefully acknowledges the financial
support of M.U.R.S.T. (within the Project "Complex problems in
Statiscal Mechanics and Field Theory") and I.N.F.N. (Italy),
as well as Prof. Arimondo and his group for useful discussions.	
%
%%%%%%%%%%%%%%%%%%%%%%%%%%%%%%%%%%%%%%%%%%%%%%%%%%%%%%%%%%%%%%%%%%
\begin{appendix}

\section{Derivation of fixed points}
\label{fixed}

Eqs. (\ref{FixP1}) can be simplified by noting their invariance 
under the global symmetry transformation
$z_{\ell} \to z_{\ell} \exp (i \Phi)$ and the fact that
$z_{\ell} /Z \in {\bf R}$ (whenever $Z\ne 0$) with ${\ell} =1,2,3$.
Then one can set $z_{\ell} \equiv x_{\ell} \, \exp(i \Phi)$,
where the $x_{\ell}$ are real numbers and $\Phi$ is an arbitrary
phase, thus reducing Eqs. (\ref{FixP1})
to a system of the three real equations 
\be
E_j({\bf x}) := \left (2U x_j^2 -\mu+
\frac{_{T}}{^2}
\right ) x_j - \frac{_{T}}{^2} X \equiv 0 \, ,
\label{FixP2}
\ee
with $j=1,2,3 $ and $X := x_1 + x_2 + x_3$.
When the condition $x_i\neq 0$ for $i=1,2,3$ is imposed, 
the latter equations can be recast in terms of an equivalent system 
of three equations one of which fixes the
Lagrange multiplier $\mu$, while the other two, now formulated in a
$\mu$-independent form, determine $x_1$, $x_2$, $x_3$ thanks to
the further condition $N = const$. In fact, the sum of 
the quantities $E_j({\bf x}) /x_j$ can be set equal to 
zero provided $x_j \ne 0$ thus giving the equation
\be
3 \mu = 2UN -
\frac{_{T}}{^2}
%\left [
\, \Sigma_i \frac{X -x_i }{x_i}
%\frac{x_2+x_3}{x_1} +\frac{x_3+x_1}{x_2} +\frac{x_1+x_2}{x_3} \right ]
\, .
\label{MU}
\ee
Moreover, from $E_1({\bf x}) /x_1 -E_2({\bf x}) /x_2 =0$ and
$E_3({\bf x}) /x_3 -E_2({\bf x}) /x_2 =0$ one obtains
\be
0=(x_2-x_1) \left [ 2U(x_2+x_1) +\frac{T\, X}{2 x_1 x_2} \right ] \, ,
\label{FixP3}
\ee
\be
0=(x_2-x_3) \left [ 2U(x_2+x_3) +\frac{T\, X}{2 x_3 x_2} \right ] \, ,
\label{FixP4}
\ee
completed by the condition $N =x_1^2+x_2^2+x_3^2$.

%%%%%%%%%%%%%%%%%%%%%%%%%%%%%%%%%%%%%%%%%%%%%%%%%%%%%%%%
\section{Derivation of fixed points of dimer type}
\label{dimerFP}

\noindent
Because of the identification
$x_1 =x_2$ (characterizing dimerlike fixed points), 
Eqs. (\ref{FixP3}) and (\ref{FixP4})
become a unique equation that can be written as
\be
2U (x^2_1 -x^2_3) = \frac{T}{2} \left [\, 1+ \frac{x_3}{x_1} -
2 \frac{x_1}{x_3}  \, \right ]\, .
\label{FPTQ}
\ee
Such an expression suggests two possible ways to parametrize
$x_1$, $x_3$
\be
\begin{array}{ll}
 x_1 = \sqrt{N} R \, {\rm ch} \, \alpha \, ,  &
x_3 = \sqrt{N} R \,{\rm sh}\, \alpha \; , 
\\
& \\
 x_1 = \sqrt{N} R \, {\rm sh}\, \beta \, , &
x_3 = \sqrt{N} R \, {\rm ch}\, \beta  
\; ,
\end{array}
\label{PAR}
\ee
both allowing for the elimination of $N$ in the constraint
$ 2 x_1^2 + x_3^2 \equiv N$. They also provide two independent
class of solutions that make explicit the three roots involved by
the cubic character of Eq. (\ref{FPTQ}).
In the first case one finds
\be
R^2 = \frac{1-q^2}{2+q^2} \, , \quad R^2 = \frac{\tau}{4}
\left [ 1 + q -
\frac{2}{q}  \right ] \, ,
\label{EQ}
\ee
[the first formula comes from the constraint on the total
number of bosons, the second one comes from Eq. (\ref{FPTQ})],
whereas the second choice gives
\be
R^2 = \frac{1-p^2}{1+ 2p^2}\, , \quad R^2 = \frac{\tau}{4}
\left [\, 2\, p - 1 -
\frac{1}{p}  \right ] \, ,
\label{EP}
\ee
where $\tau := T/UN$, $q= {\rm th} \alpha$,
$p= {\rm th} \beta$, and $\alpha,\, \beta \in {\bf R}$.
System (\ref{EQ}) reduces to the cubic equation
\be
\tau (2+q^2)(2+q) + 4q\, (1+q) =0 \, ,
\label{CUBQ}
\ee
which, provided $q \in [-1, +1]$ in order to ensure the condition
$R^2 \geq 0$, supplies either two or none solutions, depending on
the fact that $\tau < \tau_*$, $\tau > \tau_*$.
By solving the system one finds that the two roots $q_{\nu}(\tau)$,
$\nu =1,2 $ range in $[-1, 0]$ and fulfil the conditions
$$
-1\le q_1 \le q_2 \le 0 \, , \quad  {\rm for} \, \,\, 
0 \le \tau \le \tau^* \, ,
$$
with $q_{1}(\tau) = q_{2}(\tau)$ for $\tau =\tau^*$.
The parameter $\tau_*$ is identified by imposing,
in addition to Eq. (\ref{CUBQ}), the requirement that
the two $q$-dependent functions of Eqs. (\ref{EQ}) (that is,
the two right-hand sides) are tangent at some point
\be
\frac{-6q}{(2+q^2)^2} \equiv \frac{\tau}{4}
\left [ 1 + \frac{2}{q^2}  \right ]\, .
\label{Par}
\ee
Equations (\ref{CUBQ}) and (\ref{Par}), solved numerically,  
supply the value $\tau^* \simeq 0.29718$. 
 
Such a structure does not characterize system (\ref{EP}),
which always exhibits a single solution for some appropriate
value in the sector $p \in [-1/2, 0]$ obtained
from the equation
\be
\tau (1+2 p^2)(1+ 2p) + 4p\, (1+ p) =0 \, .
\label{CUBP}
\ee
In view of the restriction $q, \, p \,  < 0$, from
definitions (\ref{PAR}) one deduces that the fixed-point
coordinates are such that $x_1 x_3 < 0$.
The three solutions $q_1(\tau)$, $q_2(\tau)$, $p(\tau)$ just
obtained correspond, within the space of coordinates
$\{ (x_1, x_2, x_3) \} \equiv {\bf R}^3$, to three vectors 
expressed as
\bq
\begin{array}{c}
A_1 := \left (
a_1 , a_1, -(a_1/|a_1|) \sqrt{N-2 a_1^2}
\right ), \, \cr
{\-}\cr
A_2 := \left (
a_2 , a_2, -(a_2/|a_2|)\sqrt{N-2 a_2^2}
\right ), \, \cr
{\-}\cr
A_3 := \left (
a_3 , a_3, -(a_3/|a_3|)\sqrt{N-2 a_3^2}
\right ), \,
\end{array}
\label{firstcase}
\eq
where
\be
a_3 =\pm
%\sqrt{ \frac{N p^2}{1 +2p^2}  }
\left [ \frac{N p^2}{1 +2p^2} \right ]^{\frac{1}{2}}
%[Np^2/(1 +2p^2)]^{1/2},
\, , \quad
a_\nu= \pm
%\sqrt{ \frac{N}{2 +q^2_\nu} }
\left [ \frac{N}{2 +q^2_\nu} \right ]^{\frac{1}{2}} ,
%[N/(2 +q^2_\nu)]^{1/2},
\label{param} 
\ee
%with $\nu=1, 2$, and
with $q_\nu$, $p$ solving Eqs. (\ref{CUBQ}), (\ref{CUBP}).
One can easily check that the two $\tau$-dependent curves
$A_1$ and $A_2$ can be seen as two branches of a unique curve
based at the common point $A_1(q_1) = A_2(q_2)$ for
$\tau = \tau^*$, where they join smoothly.
%$-1/2 \le q_2(\tau) < q(\tau^*) < q_1(\tau) \le 0$
%for any given value $\tau < \tau^*$.
%%%%%%%%%%%%%%%%%%%%%%%%%%%%%%%%%%%%%%%%%%%%%%%%%%%%%%%%%

Dimerlike fixed points become nine
when considering the further set of points generated
by index permutations
\be
\begin{array}{c}
B_1 := (a_1, -(a_1/|a_1|)\sqrt{N-2 a_1^2}, a_1), \, \\
{\-}\\
B_2 := (a_2, -(a_2/|a_2|)\sqrt{N-2 a_2^2}, a_2), \, \\
{\-}\\
B_3 := (a_3, -(a_3/|a_3|)\sqrt{N-2 a_3^2}, a_3), \,
\end{array}
\ee
\be
\begin{array}{c}
C_1 := (-(a_1/|a_1|)\sqrt{N-2 a_1^2}, a_1 ,  a_1), \, \\
{\-}\\
C_2 := (-(a_2/|a_2|)\sqrt{N-2 a_2^2}, a_2 , a_2), \, \\
{\-}\\
C_3 := (-(a_3/|a_3|\sqrt{N-2 a_3^2}, a_3 , a_3), \,
\end{array}
\ee
related to the subcases
$x_1 =$ $x_3 \ne$ $ x_2$ and $x_3 =$ $x_2 \ne$ $ x_1$.
Such nine curves become actually six. Our previous
observation on considering $A_1$ and $A_2$ as a unique
curve, in fact, readily extends the other curves
$B_1$, $B_2$, and  $C_1$, $C_2$. Owing to the double choice
$\pm$ in formula (\ref{PARF}) such curves are twelve.

In order to visualize the dimerlike fixed points,
one can plot their position vectors $A_\ell, B_\ell, C_\ell$
$\ell=1,2,3$ within the three-dimensional space by varying
their parameters $a_\ell$ ($\ell=1,2,3$) in the appropriate
range.  As noted above, in the case $\tau < \tau^*$, for $q$
ranging in the intervall $[-1,0 \, ]$,  Eq. (\ref{CUBQ}) exhibits two
solutions, whereas, for any $\tau$, Eq. (\ref{CUBP}) admits one
solution with $p\in [-1/2, 0\, ]$. The corresponding ranges of
variation for $a_\nu$ $(\nu=1,2)$ and $a_3$ are 
$$
\sqrt{\frac{N}{3} } \leq |a_\nu| \leq \sqrt{\frac{N}{2} }
\, , \quad
0 \leq |a_3| \leq \sqrt{\frac{N}{6} }  \, ,
$$
respectively. 
%Fig. (\ref{sfera}) illustrates the position of dimeric fixed points
%on the sphere $\Sigma_i x_i^2 = N$ when $\tau$ is varied in
%$[0,\infty]$.
%
The representation of the position of dimeric fixed points 
on the sphere $\Sigma_i x_i^2 = N$ provides arcs that never
intersect the one with the other when
$\tau$ ranges in $[0,\infty]$.

The double sign $\pm$ in parametrizations (\ref{param}) entails that
each curve $P_j$ ($P=A,B,C$) is formed by two disjoint curves.
For $P_3$ ($P=A,B,C$) one has
$$
a_3 \in \, [-\sqrt{N/6},0 [ \,\, , \quad a_3 \in \, ]0,\sqrt{N/6}],
$$
%since $a_3(0^\pm)=\mp\sqrt{N}$.
while for $P_1$ and $P_2$ ($P=A,B,C$) the two disjoint branches
are originated by the mappings $a_\nu \to P_\nu$ with 
$$
a_\nu\in [-\sqrt{{N/2}},-\sqrt{N/3}]\, ,\quad
a_\nu\in  [\sqrt{N/3},\sqrt{N/2}]\, .
$$
%%%%%%%%%%%%%%%%%%%%%%%%%%%%%%%%%%%%%%%%%%%%%%%%%%%%%%%
\section{}
\label{zeroeig}

\noindent
The diagonalization process of $M_p$ allows one
to prove that one of the eigenvalues is always zero.
>From the standard condition ${\rm det}(M_p -\lambda) =0$
one obtains the eigenvalue equation (upon introducing 
$\Lambda := 2\lambda/T$)
$$
\Lambda^3 + (\Sigma_j \delta_j) \Lambda^2
+(\delta_1 \delta_2 +\delta_2 \delta_3 +
\delta_3 \delta_1 -3) \Lambda +
$$
\be
+ \delta_1 \delta_2 \delta_3 +2- \Sigma_i \delta_i
=0
\label{EV1}
\ee
where $\delta_1 \delta_2 \delta_3 +2- \Sigma_j \delta_j$,
upon setting $\delta_j \equiv -(x_{\ell}+ x_k)/x_j$
owing to Eqs. (\ref{FixP2}), 
can be shown to vanish in virtue of the identity
$$
\Pi^*_j \; \frac{x_{\ell}+ x_k}{x_j} =
2\, + \,\Sigma^*_j \, \frac{x_{\ell}+ x_k}{x_j} \, .
$$
The superscript symbol $*$ recalls that the indices
${\ell}$, $k$, and $j$ must differ the one from the other.
Hence, as a general result, the diagonalization of $M_p$
entails the presence of a zero eigenvalue consistent with
the analysis of dynamics in the reduced phase space
we develop in the sequel. In view of the invariance of
the matrix trace, one also finds
$$
\Sigma^3_{j=1} \lambda_j =
-\frac{_T}{^2} \, \Sigma^3_{j=1} \delta_j
\quad \to 
$$
$$
\to \quad
\lambda_1 + \lambda_2 \equiv
\frac{_T}{^2} \,
\Sigma_j \, \frac{x_{\ell}+ x_k}{x_j} \; (\equiv -3\mu + 2UN )
$$
while the two roots
\be
\lambda_{\nu} = \frac{_T}{^4} \left [ -\Delta \pm \sqrt{ (\Delta)^2
- 4(\Delta_{123} -3)}  \right]
\label{lamp}
\ee
where, $\Delta = \Sigma_i \, \delta_i$ and 
$\Delta_{123} : =\Sigma_i \, \delta_1 \delta_2 \delta_3 / \delta_i$,
and $\nu =1, 2$, [ $\lambda_1$ ($\lambda_2$) is associated
with $-$ ($+$) in the formula], are derived from the quadratic
equation that emerges from Eq. (\ref{EV1}) when removing a
factor $\Lambda$.

%%%%%%%%%%%%%%%%%%%%%%%%%%%%%%%%%%%%%%%%%%%%%%%%%%%%%%%%%%%%%%%%
\section{}
\label{charact}

This appendix is devoted to recognize, case by case, the
minima, the maxima and the hyperbolic points within the four
(class of) states identified as fixed points.

Concerning the eigenvalues of matrix $M_q$ one must take into 
account the restriction on the displacements $\xi_i$ from $v_i$
induced by the constraint $N = \Sigma_i |z_i|^2 = const$.
After recalling that the phase of $v_i$ can be absorbed,
for each $i$, by $\xi_i$ due to its arbitrariness,
the substitution $v_i \to x_i$ implies that
$$
N = \Sigma_i  \, |\xi_i + x_i|^2  = 
\Sigma_i \left [ |\xi_i|^2  + 2 q_i x_i + x_i^2 \right] \, ,
$$
which, in turn, entails $\Sigma_i |\xi_i|^2  + 2 q_i x_i = 0 $,
namely --to first order-- the plane equation
$$
%\Sigma_i  q_i x_i = 0 \, .
q_1 x_1 + q_2 x_2 +q_3 x_3 = 0\, .
$$
It represents the restriction on the displacements that
variables $\xi_i$'s are allowed to effect.
Substituting $q_i$ with $q_i = -(x_r q_r + x_s q_s)/x_i$,
where $r, s \ne i$ (and the choice of $i$ depends in general
on the condition $x_i \ne 0$) finally gives
$$
h(q\, ; 6U, T, \mu)  \equiv
\quad \quad \quad \quad \quad \quad \quad \quad \quad
$$
$$
\left [ 2 \frac{x_r x_s}{x_i^2} (6U x_i^2 - \mu)
+\frac{T (x_r +x_s -x_i)}{x_i}  \right ] q_r q_s +
$$
\be
+ \Sigma_{j \ne i}
\left [  12U x_j^2 - \frac{\mu}{x_i^2} (x_i^2 +x_j^2)
+T \frac{x_j}{x_i} \right ] q_j^2 , 
\label{MAT}
\ee

%%%%%%%%%%%%%%%%%%%%%%%%%%%%%%%%%%%%%%%%%%%%%%%%%%%%%%%%%%%%%%%%%%%%%%%
\subsection{\, Ground-state case}

\noindent
These fixed points are characterized by the fact
that $x_j=\pm \sqrt{N/3}$, for $j=1,2,3$, and $\mu=2 UN/3 -T$.
By inserting this solutions in Eq. (\ref{H3hh}) one obtains
the Hamiltonian written as
\bq
{\cal H}_3 = E_{gs} &+& \sum_{j=1}^3 \Bigl[ 
( \frac{4}{3}UN + T )q_j^2 + T  p_j^2 \Bigr] +
\non \\
&-&\frac{T}{2} \sum_{i \neq j=1}^3 \left ( p_i p_j + q_i q_j  \right )\, ,
\non 
\eq
%\[
%{\cal H}_3 = E_{gs} + \sum_{j=1}^3 [ ( \frac{4}{3}UN + T )q_j^2 ] -T
%\left ( q_1 q_2 + q_2 q_3 +  q_3 q_1 \right ) + T \sum_{i=1}^3 p_i^2 
%-T \sum_{i<j=1}^3 p_i p_j \, ,
%\]
where $E_{gs}$ is the ground state energy defined previously.
Then, by taking the constraint $N=\Sigma_{j=1}^3 |z_j|^2$ into
account, one obtains
$\Sigma_{j=1}^3 [q_j^2+p_j^2 \pm 2 \sqrt{N/3} q_j]=0$.
For little displacements from $q_j= 0 = p_j$ the latter equation
reduces to $\Sigma_{j=1}^3q_j=0$, which implies that
\bq
{\cal H}_3 \simeq E_{gs} &+& (\frac{8}{3} U N + 3 T) 
(q_1^2 + q_2^2 + q_1 q_2) + \non \\
&&T \sum_{i=1}^3 p_i^2 
-\frac{T}{2} \sum_{i \neq j=1}^3 p_i p_j \, .
\label{H3ngs}
\eq
The eigenvalues of the Hessian
corresponding to the $q$-dependent and the 
$p$-dependent part of ${\cal H}_3$ are
$\{(3 \tau +8/3) UN/2,(9 \tau + 8)UN/2\}$
and $\{0,3 T/2,3 T/2\}$, respectively.
They are positive, coherent with the fact this is a minimum.

%%%%%%%%%%%%%%%%%%%%%%%%%%%%%%%%%%%%%%%%%%%%%%%%%%%%%%%%%%%%%%%%%%%%%%%
\subsection{\, Vortex case}

\noindent
The conditions
$|x_i|^2 = N/3$ and $\mu = (4NU + 3T)/6$
characterize vortex configurations. One thus finds
$$
h(p; 2U, T, \mu) = -T(p_1+ p_2+ p_3)^2/2 < 0 \, ,
$$
whereas from
Eq. (\ref{MAT}) one gets
$$
h(q\, ; 6U, T, \mu)  = 8UN (q^2_1 + q_2^2 +  q_1 q_2)/3 \, > 0
$$
whose eigenvalues are always postive. Vortex configurations are 
therefore saddle points.

%%%%%%%%%%%%%%%%%%%%%%%%%%%%%%%%%%%%%%%%%%%
\subsection{SDW case}

\noindent
The previous analysis shows that the (fixed point) configurations
in which one of the three well is depleted (e. g., well $i =2$) is
such that $x_2 =0$, $x_1 = -x_3= \pm \sqrt{N/2}$
($\pi$-state structure), and $ \mu = NU+T/2$.
Site index permutations allows one to obtain two 
further, similar cases. In these points, Hamiltionan (\ref{H3hh}) 
can be written as
\bq
{\cal H}_3 = E_{dw} &+& UN \Bigl[2 (q_1^2+q_3^2) -q_2^2 \Bigr] +
\non \\
&-& \frac{T}{2} \Bigl[
(\sum_{i=1}^3 q_i)^2 + (\sum_{i=1}^3 p_i)^2
\Bigr] \; ,\non
\eq
In this case, the constraint on the total
number of particles supplies the constraint
$\sum_{j=1}^3 (p_j^2 +q_j^2) \pm \sqrt{2N} (q_1-q_3) =0$, 
which reduces to $q_1-q_3 \simeq 0$ when $q_j, p_j\simeq 0$.
Then ${\cal H}_3$ takes the form
\bq
{\cal H}_3 \simeq E_{dw} + UN (4 q_1^2 -q_2^2) 
- UN p_2^2 + \non \\
- \frac{T}{2} \Bigl[ (2 q_1+q_2)^2 
+ (p_1+p_2+p_3)^2 \Bigl] \, ,
\label{H3nds}
\eq
whose Hessian matrix is endowed with the eigenvalues 
$\{(6-5\tau \pm \sqrt{5(20-12\tau +5\tau^2)}) UN/4\}$ and 
$\{0,-(2+3\tau\pm\sqrt{4-4\tau+9\tau^2})UN/4\}$ for the
$q$-dependent part and the $p$-dependent part, respectively. 
The analysis of the signature of such eigenvalues leads to identify 
the fixed points of the empty-well case with saddle points.

%%%%%%%%%%%%%%%%%%%%%%%%%%%%%%%%%%%%%%%%%%%
\subsection{\, Dimeric case}

In the dimeric case the conditions on the coordinates are
$x:=x_1=x_2\neq x_3 =:y$ (none vanishing). 
Furthermore, one has to impose the condition 
$(2 U x_j^2 -\mu)x_j= T/2\sum_{k\neq j}x_k$ 
on the Lagrange multiplier $\mu$ that becomes
\be
\mu =(2 U N -T (1+y/x+x/y))/3 \ .
\label{mu4vtx}
\ee 
Let us start by analyzing the Hessian eigenvalues of Hamiltonian $p$ 
sector namely of Eq. (\ref{hp}). 
These eigenvalues are given by Eq. (\ref{lamp}) and one has that
whenever the term $\sigma :=(\delta_1 \delta_2 +\delta_2 \delta_3 +
\delta_3 \delta_1 -3)$ is greater then zero, these eingevalues are negative.
To verify the condition $\sigma > 0$ one can proceed in the following way. 
First, one substitutes in $\sigma$ the $\delta_j$ written 
in term of thier definitions; namely in term of  
$\mu$, $T$, $U$ and the fixed points coordinates $x,y$, Second, one
writes $\mu$ in terms of $x$, $y$ and  etc. [Eq. (\ref{mu4vtx})].
Finally, one eliminates the dependence on $x$, $y$ and $T$ in $\sigma$
by choosing either the parametrization 
$$
y/x=q \, , \quad x=\pm \sqrt{N/(2+q^2)} 
$$
combined with Eq. (\ref{CUBQ}) to express the $q$-dependence of $\tau$, 
or the parametrization
$$
x/y=p \, , \quad x=\pm \sqrt{N p^2/(1+2 p^2)} 
$$
combined with equation (\ref{CUBP}) to express the $p$-dependence of
$\tau$. The expression for $\sigma$ achieved in such a way depends
on $q$ or $p$, respectively. One can show that both the expressions
are always positive in the range of definition of $q$ ($[-1,0]$) and
$p$ ($[-1/2,0]$), which means that eigenvalues (\ref{lamp}) are
all negative.
  
As usual, for working out the Hessian eigenvalues of Hamiltonian 
(\ref{hq}) related to the $q$ part of the original one, it is
necessary to take into account the constraint
$\Sigma_i |z_i|^2 = N = const$.
The latter, in the present case, becomes $q_3 \equiv -(q_1+q_2)x/y$.
By means of this condition, one can reduce the dimension of the 
eigenvalues problem related to Hamiltonian (\ref{hq}) from 9 to 4. 
The Hamiltonian $h_r(q_1,q_2,U,\mu,T)$ thereby obtained can be
further simplified through the substitutions $\mu \to \mu = [2 U N 
-T (1+y/x+2x/y)]/3$ and $y/x=q$ or $x/y=p$, depending on
the parametrization one adopts.
With the first choice ($y/x=q$), and relying on Eq. (\ref{CUBQ}), one 
finds two Hessian eigevalues one of which is always positive, whereas 
the other has an illdefined sign in the domain $q\in[-1,0]$.
By using the second parametrization $x/y=p$ and Eq. (\ref{CUBP}),
both the eigenvalues thus obtained can be proven to be negative for 
$p\in[-1/2,0]$.
In summary, in the dimeric case, for $0<\tau<\tau^*$, 
one has two saddle points and one maximum point; for $\tau^*<\tau$, 
instead, fixed points reduce to  a single maximum point.

%%%%%%%%\end{appendix}
%%%%%%%%%%%%%%%%%%%%%%%%%%%%%%%%%%%%%%%%%%%%%%%%%%%%%%%%%%%%%%%%%%%%%
%\begin{appendix}

\section{}
\label{fprc}

\noindent
The fixed-point configurations corresponding to
the 
change of Hamiltonian variables
$z_i$, $z^*_i$ $\to$ $\xi_a$ $\phi_a$, $N$, $\psi$
are obtained from the equations of motion rewritten in terms of
the new variables (see below).
Coordinates transformation (\ref{chart}) can exhibit (isolated) 
singular points in which they are not invertible.
Dimer configurations in the presence of an empty well 
provides an explicit example where
transformations (\ref{chart}) are illdefined. 
In fact, fixed points
$$
(z_i)=(\sqrt{N/2}\exp i\phi, 0,\sqrt{N/2}\exp i(\phi+\pi))
$$
correspond to the set $\{(0,0,\chi,\pi-\chi)| \forall \chi\}$,
in the new description. As regards dynamical applications, 
fortunately, this problem is bypassed because the trajectories
chosen in proximity of the periodic orbits associated to this 
kind of fixed points (as well as the PS used to study the
dynamics near the same fixed points) do not contain,
by construction, such pathological points.
Upon setting $\phi_{12} := \phi_{1} + \phi_{2}$, $s := UNt$,
the equations of motion on the {\it reduced} phase space are
given by
%
%\end{multicols}
%%%%%%%%%%%%%%%%%%%
%
$$
\frac{d\phi_1}{ds} =  1-2\xi_1+\xi_2 \, +
%\displaystyle
\frac{\tau}{2} 
%\left(
\Bigl [
\sqrt{\frac{1+\xi_2}{\xi_1-\xi_2}}\, \cos \phi_2 \, +
$$
\be
\frac{(1-2\xi_1+\xi_2)\, \cos \phi_1 }
{{\sqrt{(1-\xi_1)(\xi_1-\xi_2)}}}	
	-\sqrt{\frac{1+\xi_2}{1-\xi_1}}\, \cos \phi_{12}
%\right )
\Bigr ]
\ee
%%%%%%%%%%%%%%%%%%%%%%%%%%%%%%%%%%%%%%%%%%%%%%%%
$$
-\frac{d\phi_2}{ds}= 
1 - \xi_1+2 \xi_2 + \frac{\tau}{2} 
%\displaystyle
%\left (
\Bigl [ 
\sqrt{\frac{1-\xi_1}{\xi_1-\xi_2}}\, \cos \phi_1 \, +
$$
\be
\frac{(1 - \xi_1+2 \xi_2) \,
\cos \phi_2}{{\sqrt{(\xi_1-\xi_2)(1+\xi_2)}}}
- \sqrt{\frac{1-\xi_1}{1+\xi_2}}\, \cos \phi_{12}
%\right )
\Bigr ]
\ee
%%%%%%%%%%%%%%%%%%%%%%%%%%%%%%%%%%%%%%%%%%%%%%%%%%
$$
\frac{d\xi_1}{ds}
=
\tau {\sqrt{(1-\xi_1)(\xi_1-\xi_2)}} \, \sin \phi_1 \, +
\quad\quad\quad\quad\quad
$$
\be
\quad\quad\quad
\tau {\sqrt{(1-\xi_1) (1+\xi_2)}} \, \sin \phi_{12}
\ee
%%%%%%%%%%%%%%%%%%%%%%%%%%%%%%%%%%%%%%%%%%%%%%%%%%%%%%%%%%5
$$
\frac{d\xi_2}{ds}
=
\tau {\sqrt{(\xi_1 - \xi_2)(1+\xi_2)}}\, \sin \phi_2 \, +
\quad\quad\quad\quad\quad
$$
\be
\quad\quad\quad
\tau {\sqrt{(1-\xi_1) (1+\xi_2)}} \, \sin \phi_{12} \, .
\label{fp2}
\ee

\end{appendix}

%%%%%%%%%%%%%%%%%%%%%%%
%%%%%%%%%%%%%%%%%%%%%%%%%%%%%%%%%%%%%%%%%%%%%%%%%%%%%%%%%%%%
%%%%%%%%%%%%%%%%%%%%%%%%%%%%%%%%%%%
%%%%%%%%%%%%%%%%%%%%%%%%%%%%%%%%%%%%%%%%%%%%%%%%%
%%%%%%%%%%%%%%%%%%%%%%%%%%%%%%%%%%%%%%%%%%%

%\begin{multicols}{2}
%

%%%%%%%%%%%\end{multicols}
%\vfill\eject
%%%%%%%%%%%%%%%%%%%%%%%%%%%%%%%%%%%%%%%%%%%%%%%%%%%%%%%%%%%%%%%%%%%%%%%%%
%%%%%%%%%%%%%%%%%%%% FIGURE CAPTIONS %%%%%%%%%%%%%%%%%%%%%%%%%%%%%%%%%%%%
%
%\noindent \begin{figure}[htbp]
%\includegraphics[width=7.5cm]{sfera.ps}
%\caption{Graph illustrating the dimeric fixed point location on the
%unitary sphere.}  \label{sfera}  \end{figure}
%%%%%%%%%%%%%%%%%%%%%%%% GS FIGURE %%%%%%%%%%%%%%%%%%%%%%%%%%%%%%%%%%%%%%

\noindent
\begin{figure}[htbp]
\caption{Poincar\'e section at $E\simeq 0.47 (UN^2/2)$ and
$\xi_2=-1/3$. In the neighbourhood of the ground-state
the phase-space trajectories are regular.}
\label{secgs}
\end{figure}

%\noindent \begin{figure}[htbp]
%%%\includegraphics[width=6.cm,angle=-90]{lyapgs.ps}
%\caption{Lyapunov exponent vs. time of same (regular) trajectories
%choosen close to the ground-state periodic orbit. The time decay
%confirm the stability of such trajectories.}
%\label{lyapgs} \end{figure}

\noindent
\begin{figure}[htbp]
\caption{Tipical time evolution of the condensates' populations related to
a motion with initial conditions close to a ground-state configuration.
Solid line refers to $n_1(t)$, dashed line corresponds to $n_2(t)$,
$n_3(t)$ is represented by the dotted line. The dynamics appears to be 
periodic (within the simulation time scale).}
\label{popgs}
\end{figure}

%%%%%%%%%%%%%%%%%%%%%%%%%%% VORTEX FIGURE %%%%%%%%%%%%%%%%%%%%%%%%%%%%%%%%%%%%%
\noindent
\begin{figure}[htbp]
\caption{Poincar\'e section at $E\simeq 0.77 (UN^2/2)$ and
$\xi_2=-0.3$ for orbits close to a vortexlike fixed point.
Even if some regular orbits are present, phase-space
trajectories show a dominating chaotic character.}
\label{secvrtx}
\end{figure}

%\noindent \begin{figure}[htbp]
%%%\includegraphics[width=6.cm,angle=-90]{lypregchavrtx.ps}
%\caption{The solid line traces the Lyapunov exponent as a function of time for 
%the regular trajectory of Fig. {\protect{ \ref{secregvrtx}}}; the dashed
%line show the Lyapunov exponent vs. time of the chaotic orbit of 
%Fig. {\protect{ \ref{secchavrtx}}}.} \label{lypregchavrtx} \end{figure}

\noindent
\begin{figure}[htbp]
\caption{Representation of a regular orbit (identified by
the almost continuous line) and of a chaotic one chosen
among those of the PS
%Poincar\'e section
of Fig. {\protect{\ref{secvrtx}}}.}
\label{secregchavrtx}
\end{figure}

%\noindent \begin{figure}[htbp]
%%%%\includegraphics[width=6.cm,angle=-90]{secregvrtx.ps}
%\caption{Representation of a regular trajectory of the Poincar\'e 
%section illustrated in Fig. {\protect{\ref{secvrtx}}}.}
%\label{secregvrtx}  \end{figure}

\noindent
\begin{figure}[htbp]
\caption{Time evolution of condensate populations
$n_1(t)$ (solid line),
$n_2(t)$ (dashed line) and $n_3(t)$ (dotted line)
for the regular trajectory
of Fig. {\protect{\ref{secregchavrtx}}}.}
\label{popregvrtx}
\end{figure}

%\noindent \begin{figure}[htbp]
%\includegraphics[width=6.cm,angle=-90]{secchavrtx.ps}
%\caption{Representation of a chaotic trajectory of the Poincar\'e 
%section illustrated in Fig. {\protect{\ref{secvrtx}}}.}
%\label{secchavrtx} \end{figure}

\noindent
\begin{figure}[htbp]
\caption{Populations $n_1$ (solid line),
$n_2$ (dashed line) and $n_3$ (dotted line)
are plotted as a function of the time for the chaotic trajectory 
showed in Fig. {\protect{\ref{secregchavrtx}}}.}
\label{popchavrtx}
\end{figure}

%%%%%%%%%%%%%%%%% SDW= ONE EMPTY WELL FIGURE %%%%%%%%%%%%%%%%%%%%%%%%%%%
\noindent
\begin{figure}[htbp]
\caption{Poincar\'e section at
$E\simeq 1.09 (UN^2/2)$ and $\xi_2=-0.002$ 
relative to orbits close to a SDW fixed point}
\label{secoew}
\end{figure}

\noindent
\begin{figure}[htbp]
\caption{
Representation of
one of the chaotic trajectories contained in the Poincar\'e 
section of Fig.  {\protect{\ref{secoew}}}.}
\label{secchaoew}
\end{figure}

\noindent
\begin{figure}[htbp]
\caption{
The figure plots the populations $n_1 (t)$ (solid line),
$n_2 (t)$ (dashed line) and $n_3 (t)$ (dotted line),
as a function of the time, for the chaotic trajectory showed in Fig. 
{\protect{\ref{secchaoew}}}.}
\label{popchaoew}
\end{figure}

%\noindent \begin{figure}[htbp]
%\includegraphics[width=6.cm,angle=-90]{secregoew.ps}
%\caption{One of the regular trajectories of those of the Poincar\'e 
%section illustrates in Fig.  {\protect{\ref{secoew}}}.}
%\label{secregoew} \end{figure}

\noindent
\begin{figure}[htbp]
\caption{Populations of the three condensates: $n_1$ (solid line),
$n_2$ (dashed line) and $n_3$ (dotted line)
as a function of the time for a regular trajectory of the zoom
showed in Fig. {\protect{\ref{secoew}}}.}
%Fig. {\protect{\ref{secregoew}}}.}
\label{popregoew}
\end{figure}

%\noindent \begin{figure}[htbp]
%%%\includegraphics[width=6.cm,angle=-90]{lypregchaoew.ps}
%\caption{The solid line refers to the Lyapunov exponent as a function 
%of time for the regular trajectory of Fig. {\protect{ \ref{secregoew}}}; 
%the dashed line show the Lyapunov exponent vs. time of the chaotic orbit of 
%Fig. {\protect{ \ref{secchaoew}}}.}
%\label{lypregchaoew} \end{figure}

%%%%%%%%%%%%%%%%%%%%%%%%%%%Figure sella1 %%%%%%%%%%%%%%%%%%%%%%%%

\noindent
\begin{figure}[htbp]
\caption{Poincar\'e section firmed by the conditions $\xi_2=-0.295$ 
and $E\simeq 0.73 (UN^2/2)$, close to a dimeric (saddle) fixed
point.}
\label{secsella1}
\end{figure}

\noindent
\begin{figure}[htbp]
\caption{This figure shows a regular trajectory and a chaotic one
chosen among those of Fig. \ref{secsella1}.}
\label{secregsella1}
\end{figure}

%\noindent \begin{figure}[htbp]
%\includegraphics[width=6.cm,angle=-90]{secchasella1.ps}
%\caption{Figure shows one chaotic trajectory of those of Fig. 
%\ref{secsella1}.}
%\label{secchasella1} \end{figure}

\noindent
\begin{figure}[htbp]
\caption{Temporal behavior of the condensates populations related to
the regular trajectory of Fig. \ref{secregsella1}.
The solid line, the dashed line,
and the dotted line refer to $n_1(t)$, $n_2(t)$ and $n_3(t)$,
respectively.}
\label{popregsella1}
\end{figure}

\noindent
\begin{figure}[htbp]
\caption{Condensates populations, as a function of the time,
related to the chaotic trajectory
%of Fig. \ref{secchasella1}
of Fig. \ref{secregsella1}. The solid line, the dashed line,
and the dotted line refer to $n_1(t)$, $n_2(t)$ and $n_3(t)$,
respectively.}
\label{popchasella1}
\end{figure}

%\noindent \begin{figure}[htbp]
%\includegraphics[width=6.cm,angle=-90]{lypregchasella1.ps}
%\caption{The solid line refers to the Lyapunov exponent as a function 
%of time for the regular trajectory of Fig. {\protect{ \ref{secregsella1}}}; 
%the dashed line show the Lyapunov exponent vs. time of the chaotic orbit of 
%Fig. {\protect{ \ref{secchasella1}}}.}
%\label{lypregchasella1} \end{figure}

%%%%%%%%%%%%%%%%%%%%%%%%%%%Figure sella2 %%%%%%%%%%%%%%%%%%%%%%%%

\noindent
\begin{figure}[htbp]
\caption{Poincar\'e section firmed by the conditions $\xi_2=-0.005$ 
and $E\simeq 0.91 (UN^2/2)$, close to a saddle dimeric fixed
point.}
\label{secsella2}
\end{figure}

\noindent
\begin{figure}[htbp]
\caption{This figure shows a regular trajectory and a chaotic one
chosen among those of Fig. \ref{secsella2}.}
\label{secregsella2}
\end{figure}

%\noindent \begin{figure}[htbp]
%\includegraphics[width=6.cm,angle=-90]{secchasella2.ps}
%\caption{Figure shows one chaotic trajectory of those of Fig. 
%\ref{secsella2}.}
%\label{secchasella2} \end{figure}

\noindent
\begin{figure}[htbp]
\caption{Temporal behavior of the condensates populations related to
the regular trajectory of Fig. \ref{secregsella2}. The solid line,
the dashed line, and the dotted line trace $n_1(t)$, $n_2(t)$ and
$n_3(t)$, respectively.}
\label{popregsella2}
\end{figure}

\noindent
\begin{figure}[htbp]
\caption{Condensates populations, as a function of the time,
related to the chaotic trajectory of Fig. \ref{secregsella2}.
%Fig. \ref{secchasella2}.
The solid line, the dashed line, and the dotted line trace
$n_1(t)$, $n_2(t)$ and $n_3(t)$ respectively.}
\label{popchasella2}
\end{figure}

%\noindent \begin{figure}[htbp]
%\includegraphics[width=6.cm,angle=-90]{lypregchasella2.ps}
%\caption{The solid line refers to the Lyapunov exponent as a function 
%of time for the regular trajectory of Fig. {\protect{ \ref{secregsella2}}}; 
%the dashed line show the Lyapunov exponent vs. time of the chaotic orbit of 
%Fig. {\protect{ \ref{secchasella2}}}.}
%\label{lypregchasella2} \end{figure}

%%%%%%%%%%%%%%%%%%%%%%%%%%%Figure MAX %%%%%%%%%%%%%%%%%%%%%%%%

\noindent
\begin{figure}[htbp]
\caption{
Poincar\'e section near a maximum at $E\simeq 1.7 (UN^2/2)$ and 
$\xi_2\simeq 0.8$.
In the neighbourhood of the maxima the phase space
trajectories are regular.}
\label{secMax}
\end{figure}

%\noindent \begin{figure}[htbp]
%\includegraphics[width=6.cm,angle=-90]{lyapMax.ps}
%\caption{Lyapunov exponent vs. time of same (regular) trajectories
%choosen close to a maximum periodic orbit. The time decay confirm
%the stability of such trajectories.}
%\label{lyapMax} \end{figure}

\noindent
\begin{figure}[htbp]
\caption{Time evolution of the condensates populations related to
a motion with initial conditions close to a maximum state configuration.
Figure shows $n_1(t)$ (solid line), $n_2(t)$ (dashed line), and $n_3(t)$ 
(dotted line). The motion appears to be regular.}
\label{popMax}
\end{figure}

%\end{multicols}


\begin{thebibliography}{999}
\bibitem{COM1}
Apparently, a further case should be considered, which is
described by the conditions $x_1,x_2 \ne x_3$,  $x_1\ne x_2$.
This case can be shown to have no solution.
\bibitem{COM2}
It is important to stress the fact that the coordinate change 
expressed by definitions (\ref{chart}) is well defined only for 
$z_1,z_2,z_3\neq 0$. Therefore, situations where such particular 
points are involved --fixed points of the $\pi$-state case provide 
an example of this pathology-- must be dealt with some caution.
\end{thebibliography}
\end{document}